\documentclass[12pt]{article}%
\usepackage{amsmath,latexsym}
\usepackage{graphicx}
\usepackage{amsmath}
\usepackage{amsfonts}
\usepackage{amssymb}%
\begin{document}

\title{{Exploration of traversable wormholes sustained 
   by an extra spatial dimension}}
   \author{
Peter K. F. Kuhfittig*\\  \footnote{kuhfitti@msoe.edu}
 \small Department of Mathematics, Milwaukee School of
Engineering,\\
\small Milwaukee, Wisconsin 53202-3109, USA}

\date{}
 \maketitle

\begin{abstract} The main goal in this paper is to
determine the effect of an extra dimension on a
traversable wormhole.  Here an earlier study by the
author [Phys. Rev. D \textbf{98}, 064041 (2018)] is
extended in several significant ways.  To begin with, the
extra spatial dimension is assumed to be time dependent,
while the redshift and shape functions, as well as the
extra dimension, are functions of both $r$ and $l$, the
respective radial and extra coordinates; the last of these
is therefore a function of $r$, $l$, and $t$.  The main
objective is to determine the conditions that allow the
throat of the wormhole to be threaded with ordinary matter
(by respecting the null energy condition) and that the same
conditions lead to a violation of the null energy
condition in the fifth dimension, which is therefore
responsible for sustaining the wormhole.  The dependence
of the extra dimension on $l$ and $t$ is subject to
additional conditions that are subsequently analyzed
in this paper.  Finally, the extra dimension may be
extremely small or even curled up.
\end{abstract}

\emph{Keywords:} traversable wormholes, extraspatial 
dimensions, time dependence

\section{Introduction}
While wormholes are as good a prediction of Einstein's theory
as black holes, such structures can only be maintained by
violating the null energy condition, requiring the use of
``exotic matter" \cite{MT88}.  Such wormholes would be
subject to extreme fine-tuning of the metric coefficients
in order to exist \cite{pK09}.  Another possibility for the
theoretical construction is the use of phantom dark energy,
which is known to violate the null energy condition \cite
{sS05, fL05}.

Other departures from classical relativity have been
suggested for dealing with the energy violation.  Refs. \cite
{pK18a} and \cite{pK18}, as well as the present paper,
rely on an extra spatial dimension, discussed further below.
Part of the motivation comes from Lobo and Oliveira \cite{LO09}
in the context of $f(R)$ modified gravity: in principle, a
wormhole could be constructed from ordinary matter, while the
unavoidable violation of the null energy condition can be
attributed to the effect of the modified gravitational theory.
In analogous fashion, Refs. \cite{fR12, pK15} show that
ordinary matter may be allowed since the violation of the
null energy condition is due to the noncommutative-geometry
background.

A suitable model for a five- or higher-dimensional wormhole
is suggested by the following classical line element in
Schwarzschild coordinates:
\begin{equation}\label{E:L1}
  ds^2=-e^{2\Phi(r)}dt^2 +e^{2\lambda(r)}dr^2 +r^2S_2^2,
\end{equation}
where $S_2^2=d\theta^2+\text{sin}^2\theta\,d\phi^2.$  In a
seemingly natural extension, the two-sphere $S_2^2$ is
replaced by the $n$-sphere $S_n^2$, where
\begin{equation}\label{E:product}
   dS_n^2=d\theta_1^2+\text{sin}^2\theta_1\,d\theta_2+
   \text{sin}^2\theta_1\,\text{sin}^2\theta_2\,d\theta_3
   +\cdot\cdot\cdot+\prod_{i=1}^{n-1}
   \text{sin}^2\theta_i\,d\theta^2_n
\end{equation}
\cite{KS96, CBA96}.  For a 3-sphere, we can  use
the familiar notation
\begin{equation}
  S_3^2=d\theta^2+\text{sin}^2\theta\,
  (d\phi^2+\text{sin}^2\phi\,d\alpha^2).
\end{equation}
This form arises in the classical
Friedmann-Lema\^{i}tre-Robertson-Walker (FLRW) model
\begin{equation}
  ds^2=-d\tau^2+[a(t)]^2\left[\frac{dr^2}{1-kr^2}+
  d\theta^2+\text{sin}^2\theta\,(d\phi^2
  +\text{sin}^2\phi\,d\alpha^2)\right].
\end{equation}

Wormhole solutions based on line element (\ref{E:product})
are discussed in Ref. \cite{fR12}, given a
noncommutative-geometry background.

Since Morris-Thorne wormholes do not normally require a
cosmological setting, it seems more natural to consider
the line element proposed in Ref. \cite{pK18a}:
\begin{equation}\label{E:L2}
  ds^2=-e^{2\Phi(r)}dt^2 +e^{2\lambda(r)}dr^2+r^2
  (d\theta^2+\text{sin}^2\theta\,d\phi^2)\\+e^{2\mu(r,l)}dl^2,
\end{equation}
where $l$ is the extra coordinate.  (We are using units in
which $c=G=1$.)

The idea of an extra spatial dimension had its origin in the
Kaluza-Klein theory and eventually led to the compactified
extra dimensions in string theory.  That an extra dimension
may be macroscopic was proposed by Paul Wesson in a series
of papers \cite{pW11, WP92, pW15, pW13}, leading to the
\emph{induced-matter theory:} the field equations for a
five-dimensional totally flat space yield the Einstein
field equations in four dimensions containing matter.  In
this paper, we assume that the extra dimension may be time
dependent and that no \emph{a priori} restriction is to be
placed on the size.

Of particular interest to us is the form of the line element
in Ref. \cite{pW11}:
\begin{equation}
   ds^2=(l/L^2)g_{\alpha\beta}(x^{\gamma}, l)dx^{\alpha}
   dx^{\beta}\pm dl^2.
\end{equation}
Here $dl^2$ can take on the more general form
$A(x^{\gamma}, l)dl^2$.  To emphasize the dependence of
the metric coefficients on $l$, we will employ the
following line element for a wormhole based on an extension
of Eq. (\ref{E:L2}):
\begin{equation}\label{E:main}
  ds^2=-e^{2\Phi(r,l)}dt^2 +e^{2\lambda(r,l)}dr^2+r^2
  (d\theta^2+\text{sin}^2\theta\,d\phi^2)\\+e^{2\mu(r,l,t)}dl^2.
\end{equation}
(To make our results as general as possible, we also assume
that $\mu$ may be time dependent.)

Before continuing, let us recall the basic properties of a
Morris-Thorne wormhole.  Letting $e^{2\lambda(r,l)}=
1-b(r,l)/r$, the line element becomes
\begin{equation}\label{E:L4}
  ds^2=-e^{2\Phi(r,l)}dt^2+\frac{dr^2}{1-b(r,l)/r}+r^2
  (d\theta^2+\text{sin}^2\theta\,d\phi^2)\\+e^{2\mu(r,l,t)}dl^2.
\end{equation}
Here $\Phi=\Phi(r,l)$ is called the \emph{redshift function},
which must be finite everywhere to prevent the occurrence
of an event horizon; $b=b(r,l)$ is called the \emph{shape
function}.  The spherical surface $r=r_0$ is called the
\emph{throat} of the wormhole.  The shape function must
satisfy the following conditions: $b(r_0,l)=r_0$ and
$\partial b(r_0,l)/\partial r<1$, called the
\emph{flare-out condition} in Ref. \cite{MT88}.  These
conditions can only be met by violating the null energy
condition (NEC), which states that for the energy
momentum tensor
$T_{\alpha\beta}$,
\begin{equation}\label{E:null}
   T_{\alpha\beta}k^{\alpha}k^{\beta}\ge 0\,\,
   \text{for all null vectors}\,\, k^{\alpha}.
\end{equation}
To see the significance in a wormhole setting, we assume
an orthonormal frame and note that $T_{00}=\rho$
is the energy density and $T_{11}=p_r$ is the radial
pressure.  Now the outgoing radial null vector
$(1,1,0,0)$ yields $\rho +p_r<0$ whenever the NEC
is violated.

Our new model, Eq. (\ref{E:main}), greatly extends the
discussion in Ref. \cite{pK18a}, which assumes that
the metric coefficients are independent of $l$.  According
to Ref. \cite{pW11}, the dependence on $l$ is associated
with the presence of matter, whose nature can be determined
by reducing the 5$D$ Ricci-flat equation $R_{AB}=0$ to the
4$D$ Einstein equation $G_{\alpha\beta}=8\pi T_{\alpha\beta}$.

As noted earlier, our main goal is to determine the conditions
that allow the throat of the wormhole to be threaded with
ordinary matter and that the same conditions result in a
violation of the NEC in the fifth dimension.  The conditions
discussed include the effect of the time dependence.

\section{The solution}\label{S:solution}
In this section we obtain the general solution from line
element (\ref{E:L4}) and study the consequences in subsequent
sections.  We choose an orthonormal basis
$\{e_{\hat{\alpha}}\}$ which is dual to the following
1-form basis:
\begin{equation}\label{E:oneform1}
    \theta^0=e^{\Phi(r,l)}\, dt,\quad \theta^1=
    \left[1-\frac{b(r,l)}{r}\right]^{-1/2}\,dr,\\
     \quad\theta^2=r\,d\theta, \quad
        \theta^3=r\,
   \,\text{sin}\,\theta\,d\phi,\quad \theta^4=e^{\mu(r,l,t)}dl.
\end{equation}
Alternatively,
\begin{equation}\label{E:oneform3}
     dt=e^{-\Phi(r,l)}\,\theta^0,\quad dr=
     \left[1-\frac{b(r,l)}{r}\right]^{1/2}\,\theta^1,\\
     \quad d\theta=\frac{1}{r}\theta^2, \quad
    d\phi=\frac{1}{r\,\text{sin}\,\theta}\theta^3, \quad
    dl=e^{-\mu(r,l,t)}\,\theta^4.
\end{equation}
Following Ref. \cite{HT90}, we use the method of differential
forms to obtain the connection 1-forms, the curvature 2-forms,
and the resulting components of the Riemann curvature tensor.
To that end, we calculate the exterior derivatives in terms
of the basis $\{\theta^i\}$:
\begin{equation}
   d\theta^0=\frac{\partial\Phi(r,l)}{\partial r}
   \left[1-\frac{b(r,l)}{r}\right]
   ^{1/2}\,\theta^1\wedge\theta^0+\frac{\partial\Phi(r,l)}
   {\partial l}e^{-\mu(r,l,t)}\,\,\theta^4\wedge\theta^0,
\end{equation}
\begin{equation}
   d\theta^1=\frac{1}{2r}\left[1-\frac{b(r,l)}{r}\right]^{-1}
   \frac{\partial b(r,l)}{\partial l}e^{-\mu(r,l,t)}\,\,
   \theta^4\wedge\theta^1,
 \end{equation}
\begin{equation}
    d\theta^2=\frac{1}{r}\left[1-\frac{b(r,l)}{r}\right]^{1/2}
    \theta^1\wedge\theta^2,
\end{equation}
\begin{equation}
     d\theta^3=\frac{1}{r}\left[1-\frac{b(r,l)}{r}\right]^{1/2}
     \theta^1\wedge\theta^3+\frac{1}{r}\text{cot}\,\theta
     \,\,\theta^2\wedge\theta^3,
\end{equation}
\begin{equation}
   d\theta^4=\frac{\partial \mu(r,l,t)}{\partial r}
   \left[1-\frac{b(r,l)}{r}\right]^{1/2}
   \,\theta^1\wedge\theta^4+e^{-\Phi(r,l)}
   \frac{\partial\mu(r,l,t)}{\partial t}\,\,
   \theta^0\wedge\theta^4.
\end{equation}
The connection 1-forms $\omega^i_{\phantom{i}\,\,k}$ have the
symmetry
    $\omega^0_{\phantom{i}\,\,i}=\omega^i_{\phantom{0}0}
    \;(i=1,2,3,4)$\;\text{and}\;$\omega^i_{\phantom{j}j}=
     -\omega^j_{\phantom{i}\,i}\;(i,j=1,2,3,4, i\ne j)$,
and are related to the basis $\{\theta^i\}$ by
\begin{equation}
   d\theta^i=-\omega^i_{\phantom{k}k}\wedge\theta^k.
\end{equation}
The solution of this system is
\begin{equation}
   \omega^0_{\phantom{0}1}=\frac{d\Phi(r,l)}{dr}
   \left[1-\frac{b(r,l)}{r}\right]^{1/2}\theta^0,
\end{equation}
\begin{equation}
  \omega^0_{\phantom{0}4}=\frac{\partial\Phi(r,l)}{\partial l}
  e^{-\mu(r,l,t)}\,\theta^0+e^{-\Phi(r,l)}
  \frac{\partial\mu(r,l,t)}{\partial t}\,\theta^4,
\end{equation}
\begin{equation}
   \omega^2_{\phantom{0}1}=
   \frac{1}{r}\left[1-\frac{b(r,l)}{r}\right]^{1/2}\theta^2,
\end{equation}
\begin{equation}
    \omega^3_{\phantom{0}1}=
    \frac{1}{r}\left[1-\frac{b(r,l)}{r}\right]^{1/2}\theta^3,
\end{equation}
\begin{equation}
   \omega^3_{\phantom{0}2}=
   \frac{1}{r}\,\text{cot}\,\theta\,\,\theta^3,
\end{equation}
\begin{equation}\label{E:omega}
   \omega^4_{\phantom{0}1}=
   \frac{\partial\mu(r,l,t)}{\partial r}
      \left[1-\frac{b(r,l)}{r}\right]^{1/2}\theta^4
      -\frac{1}{2r}\left[1-\frac{b(r,l)}{r}\right]^{-1}
      \frac{\partial b(r,l)}{\partial l}
      e^{-\mu(r,l,t)}\,\theta^1,
\end{equation}
\begin{equation}
   \omega^0_{\phantom{0}2}=\omega^0_{\phantom{0}3}=
   \omega^2_{\phantom{0}4}=\omega^3_{\phantom{0}4}=0.
\end{equation}

The curvature 2-forms $\Omega^i_{\phantom{j}j}$ are calculated
directly from the Cartan structural equations
\begin{equation}
    \Omega^i_{\phantom{j}j}=d\omega^i_{\phantom{j}j} +\omega^i
     _{\phantom{j}k}\wedge\omega^k_{\phantom{j}j}.
\end{equation}
and are listed next.  (To shorten the expressions, we will
denote $\Phi(r,l)$ by $\Phi$, $b(r,l)$ by $b$, and $\mu(r,l,t)$
by $\mu$.)
\begin{multline}
\Omega^0_{\phantom{0}1}=\left[-\frac{\partial^2\Phi}{\partial r^2}
   \left(1-\frac{b}{r}\right)+\frac{1}{2r^2}\frac{\partial\Phi}
   {\partial r}\left(r\frac{\partial b}{\partial r}-b\right)
   -\left(\frac{\partial\Phi}{\partial r}\right)^2\left(1-\frac{b}{r}
   \right)\right.\\\left.-\frac{1}{2r}e^{-2\mu}\frac{\partial\Phi}{\partial l}
   \frac{\partial b}{\partial l}\left(1-\frac{b}{r}\right)^{-1}\right]
   \,\,\theta^0\wedge\theta^1\\
   +e^{-\mu}\left[-\frac{\partial^2\Phi}{\partial r\partial l}
   \left(1-\frac{b}{r}\right)^{1/2}+\frac{1}{2r}\frac{\partial\Phi}
   {\partial r}\frac{\partial b}{\partial l}\left(1-\frac{b}{r}
   \right)^{-1/2}-\frac{\partial\Phi}{\partial r}
   \frac{\partial\Phi}{\partial l}\left(1-\frac{b}{r}\right)^{1/2}
   \right. \\ \left.
   +\frac{\partial\Phi}{\partial l}\frac{\partial\mu}{\partial r}
   \left(1-\frac{b}{r}\right)^{1/2}\right]\,\theta^0\wedge\theta^4
   +\frac{1}{2r}e^{-\Phi}e^{-\mu}\frac{\partial b}{\partial l}
   \frac{\partial\mu}{\partial t}\left(1-\frac{b}{r}\right)^{-1}
   \,\theta^1\wedge\theta^4,
\end{multline}
\begin{equation}
   \Omega^0_{\phantom{0}2}=-\frac{1}{r}\frac{\partial\Phi}{\partial r}
   \left(1-\frac{b}{r}\right)\theta^0\wedge\theta^2,
\end{equation}
\begin{equation}
   \Omega^0_{\phantom{0}3}=-\frac{1}{r}\frac{\partial\Phi}{\partial r}
   \left(1-\frac{b}{r}\right)\theta^0\wedge\theta^3,
\end{equation}
\begin{multline}
   \Omega^0_{\phantom{0}4}=
   e^{-\mu}\left(1-\frac{b}{r}\right)^{1/2}\left(
   -\frac{\partial^2\Phi}{\partial r\partial l}
   +\frac{\partial\Phi}{\partial l}\frac{\partial\mu}
   {\partial r}-\frac{\partial\Phi}{\partial r}
   \frac{\partial\Phi}{\partial l}\right)\,\theta^0\wedge\theta^1
   \\+\frac{1}{2r}e^{-\mu}\frac{\partial\Phi}{\partial r}
   \frac{\partial b}{\partial l}\left(1-\frac{b}{r}
   \right)^{-1/2}\,\theta^0\wedge\theta^1\\
   +\left[-e^{-2\mu}\frac{\partial^2\Phi}{\partial l^2}
   +e^{-2\mu}\frac{\partial\Phi}{\partial l}
   \frac{\partial\mu}{\partial l}-e^{-2\mu}\left(
   \frac{\partial\Phi}{\partial l}\right)^2
   +e^{-2\Phi}\frac{\partial^2\mu}{\partial t^2}+e^{-2\Phi}
   \left(\frac{\partial\mu}{\partial t}\right)^2\right.\\
   \left. -\frac{\partial\Phi}{\partial  r}\frac{\partial\mu}
   {\partial r}\left(1-\frac{b}{r}\right)\right]\,
   \theta^0\wedge\theta^4\\
   +e^{-\Phi}\left(1-\frac{b}{r}\right)^{1/2}
   \left(-\frac{\partial\Phi}{\partial r}
   \frac{\partial\mu}{\partial t}+\frac{\partial^2\mu}
   {\partial r\partial t}+\frac{\partial\mu}{\partial r}
   \frac{\partial\mu}{\partial t}\right)\,\
   \theta^1\wedge\theta^4,
\end{multline}
\begin{equation}
  \Omega^1_{\phantom{0}2}=\frac{1}{2r^3}
  \left(r\frac{\partial b}{\partial r}-b\right)\,
     \theta^1\wedge\theta^2-\frac{1}{2r^2}e^{-\mu}
     \frac{\partial b}{\partial l}\left(1-\frac{b}{r}
     \right)^{-1/2}\,\theta^2\wedge\theta^4,
\end{equation}
\begin{equation}
  \Omega^1_{\phantom{0}3}=\frac{1}{2r^3}
  \left(r\frac{\partial b}{\partial r}-b\right)\,
     \theta^1\wedge\theta^3-\frac{1}{2r^2}e^{-\mu}
     \frac{\partial b}{\partial l}\left(1-\frac{b}{r}
     \right)^{-1/2}\,\theta^3\wedge\theta^4,
\end{equation}
\begin{multline}
   \Omega^1_{\phantom{0}4}=\left[-\frac{\partial^2\mu}
   {\partial r^2}\left(1-\frac{b}{r}\right)+\frac{1}{2r^2}
   \frac{\partial\mu}{\partial r}\left(r\frac{\partial b}
   {\partial r}-b\right)-\left(\frac{\partial\mu}
   {\partial r}\right)^2\left(1-\frac{b}{r}\right)\right.
   \\\ \left.
   -\frac{3}{4r^2}e^{-2\mu}\left(1-\frac{b}{r}\right)^{-2}
   \left(\frac{\partial b}{\partial l}\right)^2\right.\\
   \left.-\frac{1}{2r}e^{-2\mu}\left(1-\frac{b}{r}\right)^{-1}
   \frac{\partial^2b}{\partial l^2}+\frac{1}{2r}e^{-2\mu}
   \left(1-\frac{b}{r}\right)^{-1}\frac{\partial b}{\partial l}
   \frac{\partial\mu}{\partial l}\right]\,\theta^1\wedge\theta^4\\
   +\left[-e^{-\Phi}\frac{\partial^2\mu}{\partial r\partial t}
   \left(1-\frac{b}{r}\right)^{1/2}-e^{-\Phi}\frac{\partial\mu}{\partial r}
   \frac{\partial\mu}{\partial t}\left(1-\frac{b}{r}\right)^{1/2}
   \right. \\ \left.
   +e^{-\Phi}\frac{\partial\Phi}{\partial r}\frac{\partial\mu}
   {\partial t}\left(1-\frac{b}{r}\right)^{1/2}\right]
   \,\theta^0\wedge\theta^4
   -\frac{1}{2r}e^{-\Phi}e^{-\mu}\frac{\partial\mu}{\partial t}
   \frac{\partial b}{\partial l}\left(1-\frac{b}{r}\right)^{-1}
   \,\theta^0\wedge\theta^1,
\end{multline}
\begin{equation}
   \Omega^2_{\phantom{0}3}=\frac{b}{r^3}\,\theta^2\wedge\theta^3,
\end{equation}
\begin{equation}
   \Omega^2_{\phantom{0}4}=-\frac{1}{r}\frac{\partial\mu}
   {\partial r}\left(1-\frac{b}{r}\right)\theta^2\wedge\theta^4
   -\frac{1}{2r^2}e^{-\mu}\left(1-\frac{b}{r}\right)^{-1/2}
   \frac{\partial b}{\partial l}\,\theta^1\wedge\theta^2,
\end{equation}
\begin{equation}
   \Omega^3_{\phantom{0}4}=-\frac{1}{r}\frac{\partial\mu}
   {\partial r}\left(1-\frac{b}{r}\right)\theta^3\wedge\theta^4
   -\frac{1}{2r^2}e^{-\mu}\left(1-\frac{b}{r}\right)^{-1/2}
   \frac{\partial b}{\partial l}\,\theta^1\wedge\theta^3.
\end{equation}

The components of the Riemann curvature tensor can be read off
directly from the form
\begin{equation}
   \Omega^i_{\phantom{j}j}=-\frac{1}{2}R_{mnj}^{\phantom{mnj}i}
    \;\theta^m\wedge\theta^n
\end{equation}
and are listed next:
\begin{multline}
   R_{011}^{\phantom{000}0}=\frac{\partial^2\Phi(r,l)}
   {\partial r^2}\left(1-\frac{b(r,l)}{r}\right)-\frac{1}{2r^2}
   \frac{\partial\Phi(r,l)}{\partial r}\left(r
   \frac{\partial b(r,l)}{\partial r}-b(r,l)\right)\\
   +\left(\frac{\partial\Phi(r,l)}{\partial r}\right)^2
   \left(1-\frac{b(r,l)}{r}\right)+\frac{1}{2r}e^{-2\mu(r,l,t)}
   \frac{\partial\Phi(r,l)}{\partial l}\left(1-\frac{b(r,l)}{r}
   \right)^{-1}\frac{\partial b(r,l)}{\partial l},
\end{multline}
\begin{equation}
   R_{022}^{\phantom{000}0}=R_{033}^{\phantom{000}0}=
   \frac{1}{r}\frac{\partial\Phi(r,l)}{\partial r}
   \left(1-\frac{b(r,l)}{r}\right),
\end{equation}
\begin{multline}
   R_{044}^{\phantom{000}0}=e^{-2\mu(r,l,t)}\left[
   \frac{\partial^2\Phi(r,l)}{\partial l^2}
   -\frac{\partial\Phi(r,l)}{\partial l}\frac{\partial\mu(r,l,t)}
   {\partial l}+\left(\frac{\partial\Phi(r,l)}{\partial l}\right)^2
   \right]\\
   -e^{-2\Phi(r,l)}\left[\frac{\partial^2\mu(r,l,t)}{\partial t^2}
   +\left(\frac{\partial\mu(r,l,t)}{\partial t}\right)^2\right]
    +\frac{\partial\Phi(r,l)}{\partial r}\frac{\partial\mu(r,l,t)}
    {\partial r}\left(1-\frac{b(r,l)}{r}\right),
\end{multline}
\begin{equation}
   R_{122}^{\phantom{000}1}=R_{133}^{\phantom{000}1}=
   -\frac{1}{2r^3}\left(r\frac{\partial b(r,l)}{\partial r}
   -b(r,l)\right),
\end{equation}
\begin{multline}
   R_{144}^{\phantom{000}1}=\frac{\partial^2\mu(r,l,t)}{\partial r^2}
   \left(1-\frac{b(r,l)}{r}\right)+\left(\frac{\partial\mu(r,l,t)}
   {\partial r}\right)^2\left(1-\frac{b(r,l)}{r}\right)\\
   -\frac{1}{2r^2}\frac{\partial\mu(r,l,t)}{\partial r}
   \left(r\frac{\partial b(r,l)}{\partial r}-b(r,l)\right)\\
   +\frac{3}{4r^2}e^{-2\mu(r,l,t)}\left(1-\frac{b(r,l)}{r}
   \right)^{-2}\left(\frac{\partial b(r,l)}{\partial l}\right)^2\\
   +\frac{1}{2r}e^{-2\mu(r,l,t)}\left(1-\frac{b(r,l)}{r}\right)^{-1}
   \frac{\partial^2 b(r,l)}{\partial l^2}\\
   -\frac{1}{2r}e^{-2\mu(r,l,t)}\left(1-\frac{b(r,l)}{r}\right)^{-1}
   \frac{\partial b(r,l)}{\partial l}
   \frac{\partial\mu(r,l,t)}{\partial l},
\end{multline}
\begin{equation}
   R_{233}^{\phantom{000}2}=-\frac{b(r,l)}{r^3},
\end{equation}
\begin{equation}
   R_{244}^{\phantom{000}2}=R_{344}^{\phantom{000}3}=
   \frac{1}{r}\frac{\partial\mu(r,l,t)}{\partial r}
      \left(1-\frac{b(r,l)}{r}\right),
\end{equation}
\begin{multline}
   R_{041}^{\phantom{000}0}=e^{-\mu(r,l,t)}
   \left(1-\frac{b(r,l)}{r}\right)^{1/2}
   \left[\frac{\partial^2\Phi(r,l)}{\partial r
   \partial l}+\frac{\partial\Phi(r,l)}{\partial r}
   \frac{\partial\Phi(r,l)}{\partial l}\right.\\ \left.
    -\frac{\partial\Phi(r,l)}{\partial l}
    \frac{\partial\mu(r,l,t)}{\partial r}\right]\\
    -\frac{1}{2r}e^{-\mu(r,l,t)}\frac{\partial\Phi(r,l)}
    {\partial r}\left(1-\frac{b(r,l)}{r}\right)^{-1/2}
    \frac{\partial b(r,l)}{\partial l},
\end{multline}
\begin{equation}
   R_{411}^{\phantom{000}0}=\frac{1}{2r}e^{-\Phi(r,l)}
   e^{-\mu(r,l,t)}\frac{\partial\mu(r,l,t)}{\partial t}
   \left(1-\frac{b(r,l)}{r}\right)^{-1}\frac{\partial b(r,l)}
   {\partial l},
   \end{equation}
\begin{multline}
   R_{044}^{\phantom{000}1}=e^{-\Phi(r,l)}\left(1-\frac{b(r,l)}{r}
   \right)^{1/2}\left[-\frac{\partial\Phi(r,l)}{\partial r}
    \frac{\partial\mu(r,l,t)}{\partial t}+\frac{\partial^2\mu(r,l,t)}
    {\partial r\partial t}\right.\\ \left.
    +\frac{\partial\mu(r,l,t)}{\partial r}
    \frac{\partial\mu(r,l,t)}{\partial t}\right],
\end{multline}
\begin{equation}
   R_{214}^{\phantom{000}2}=R_{314}^{\phantom{000}3}=
   -\frac{1}{2r^2}e^{-\mu(r,l,t)}\left(1-\frac{b(r,l)}{r}\right)
   ^{-1/2}\frac{\partial b(r,l)}{\partial l}.
\end{equation}

The last form to be derived in this section is the Ricci
tensor, which is obtained by a trace on the Riemann
curvature tensor:
\begin{equation}\label{E:trace}
   R_{ab}=R_{acb}^{\phantom{000}c}.
\end{equation}
The various components are
\begin{multline}\label{E:R00}
 R_{00}=\frac{\partial^2\Phi(r,l)}{\partial r^2}\left(1-\frac{b(r,l)}{r}\right)
 -\frac{1}{2r^2}\frac{\partial\Phi(r,l)}{\partial r}\left(r\frac{\partial b(r,l)}
 {\partial r}-b(r,l)\right)\\+\left(\frac{\partial\Phi(r,l)}{\partial r}\right)^2
 \left(1-\frac{b(r,l)}{r}\right)
 +\frac{1}{2r}e^{-2\mu(r,l,t)}\frac{\partial\Phi(r,l)}{\partial l}\left(1-\frac{b(r,l)}
 {r}\right)^{-1}\frac{\partial b(r,l)}{\partial l}\\+\frac{2}{r}\frac{\partial\Phi(r,l)}
 {\partial r}\left(1-\frac{b(r,l)}{r}\right)+\frac{\partial^2\Phi(r,l)}{\partial l^2}
 e^{-2\mu(r,l,t)}
 -\frac{\partial\Phi(r,l)}{\partial l}\frac{\partial\mu(r,l,t)}{\partial l}e^{-2\mu(r,l,t)}\\
 +\left(\frac{\partial\Phi(r,l)}{\partial l}\right)^2e^{-2\mu(r,l,t)}-e^{-2\Phi(r,l)}
 \frac{\partial^2\mu(r,l,t)}{\partial t^2}\\
 -e^{-2\Phi(r,l)}\left(\frac{\partial\mu(r,l,t)}{\partial t}\right)^2
 +\frac{\partial\Phi(r,l)}{\partial r}\frac{\partial\mu(r,l,t)}{\partial r}
 \left(1-\frac{b(r,l)}{r}\right),
\end{multline}
\begin{multline}\label{E:R11}
 R_{11}=-\frac{\partial^2\Phi(r,l)}{\partial r^2}\left(1-\frac{b(r,l)}{r}\right)
 +\frac{1}{2r^2}\frac{\partial\Phi(r,l)}{\partial r}\left(r\frac{\partial b(r,l)}
 {\partial r}-b(r,l)\right)\\-\left(\frac{\partial\Phi(r,l)}{\partial r}\right)^2
 \left(1-\frac{b(r,l)}{r}\right)
 -\frac{1}{2r}e^{-2\mu(r,l,t)}\frac{\partial\Phi(r,l)}{\partial l}
 \left(1-\frac{b(r,l)}{r}\right)^{-1}\frac{\partial b(r,l)}{\partial l}\\+\frac{1}{r^3}
 \left(r\frac{\partial b(r,l)}{\partial r}-b(r,l)\right)-\frac{\partial^2\mu(r,l,t)}
 {\partial r^2}\left(1-\frac{b(r,l)}{r}\right)\\
 -\left(\frac{\partial\mu(r,l,t)}{\partial r}\right)^2\left(1-\frac{b(r,l)}{r}\right)
 +\frac{1}{2r^2}\frac{\partial\mu(r,l,t)}{\partial r}\left(r\frac{\partial b(r,l)}
 {\partial r}-b(r,l)\right)\\
 -\frac{3}{4r^2}\left(1-\frac{b(r,l)}{r}\right)^{-2}\left(\frac{\partial b(r,l)}
 {\partial l}\right)^2e^{-2\mu(r,l,t)}\\-\frac{1}{2r}\left(1-\frac{b(r,l)}{r}\right)^{-1}
 \frac{\partial^2 b(r,l)}{\partial l^2}e^{-2\mu(r,l,t)}\\
 +\frac{1}{2r}\left(1-\frac{b(r,l)}{r}\right)^{-1}\frac{\partial b(r,l)}{\partial l}
 \frac{\partial\mu(r,l,t)}{\partial l}e^{-\mu(r,l,t)},
\end{multline}
\begin{multline}
 R_{22}=R_{33}=-\frac{1}{r}\frac{\partial\Phi(r,l)}{\partial r}\left(1-\frac{b(r,l)}{r}
 \right)+\frac{1}{2r^3}\left(r\frac{\partial b(r,l)}{\partial r}-b(r,l)\right)\\
 +\frac{b(r,l)}{r^3}-\frac{1}{r}\left(1-\frac{b(r,l)}{r}\right)
 \frac{\partial\mu(r,l,t)}{\partial r},
\end{multline}
\begin{multline}
 R_{44}=-\frac{\partial^2\Phi(r,l)}{\partial l^2}e^{-2\mu(r,l,t)}+\frac{\partial\Phi(r,l)}
 {\partial l}\frac{\partial\mu(r,l,t)}{\partial l}e^{-2\mu(r,l,t)}\\-\left(
 \frac{\partial\Phi(r,l)}{\partial l}\right)^2e^{-2\mu(r,l,t)}+e^{-2\Phi(r,l)}
 \frac{\partial^2\mu(r,l,t)}{\partial t^2}\\
 +e^{-2\Phi(r,l)}\left(\frac{\partial\mu(r,l,t)}{\partial t}\right)^2-\frac{\partial\Phi(r,l)}
 {\partial r}\frac{\partial\mu(r,l,t)}{\partial r}\left(1-\frac{b(r,l)}{r}\right)\\
 -\frac{\partial^2\mu(r,l,t)}{\partial r^2}\left(1-\frac{b(r,l)}{r}\right)
 -\left(\frac{\partial\mu(r,l,t)}{\partial r}\right)^2\left(1-\frac{b(r,l)}{r}\right)\\
 +\frac{1}{2r^2}\frac{\partial\mu(r,l,t)}{\partial r}\left(r\frac{\partial b(r,l)}{\partial r}
 -b(r,l)\right)\\
 -\frac{3}{4r^2}\left(1-\frac{b(r,l)}{r}\right)^{-2}\left(\frac{\partial b(r,l)}
 {\partial l}\right)^2e^{-2\mu(r,l,t)}\\-\frac{1}{2r}\left(1-\frac{b(r,l)}{r}\right)^{-1}
 \frac{\partial^2 b(r,l)}{\partial l^2}e^{-2\mu(r,l,t)}\\+\frac{1}{2r}\left(1-\frac{b(r,l)}{r}
 \right)^{-1}\frac{\partial b(r,l)}{\partial l}\frac{\partial\mu(r,l,t)}{\partial l}
 e^{-2\mu(r,l,t)}\\-\frac{2}{r}\left(1-\frac{b(r,l)}{r}\right)\frac{\partial\mu(r,l,t)}
 {\partial r},
\end{multline}
\begin{multline}\label{E:R01}
 R_{01}=\left(1-\frac{b(r,l)}{r}\right)^{1/2}\left[e^{-\Phi(r,l)}\frac{\partial\Phi(r,l)}
 {\partial r}\frac{\partial\mu(r,l,t)}{\partial t}-e^{-\Phi(r,l)}\frac{\partial^2\mu(r,l,t)}
 {\partial r\partial t}\right.\\ \left.-e^{-\Phi(r,l)}\frac{\partial\mu(r,l,t)}{\partial r}
 \frac{\partial\mu(r,l,t)}{\partial t}\right],
 \end{multline}
\begin{equation}\label{E:R04}
 R_{04}=\frac{1}{2r}e^{-\Phi(r,l)}\left(1-\frac{b(r,l)}{r}\right)^{-1}\frac{\partial b(r,l)}
 {\partial l}\frac{\partial\mu(r,l,t)}{\partial t}e^{-\mu(r,l,t)}.
\end{equation}

These forms will come into play in the next section.

\section{A possible curvature singularity}
In discussing the structure of a wormhole, we recall that
at the throat, we have $b(r_0,l)=r_0$ for every $l$, leading
to the question of its physical interpretation.  The problem
takes care of itself in the following sense: consider the
Ricci scalar
\begin{equation}\label{E:Ricci}
R=R^i_{\phantom{0}i}=-R_{00}+R_{11}+R_{22}+R_{33}+R_{44}.
\end{equation}
Using Eq. (\ref{E:trace}), this can be written in the
following convenient form:
\begin{equation}
  \frac{1}{2}R=-R_{011}^{\phantom{000}0}
    -R_{022}^{\phantom{000}0}
    -R_{033}^{\phantom{000}0}
    -R_{044}^{\phantom{000}0}\\
  -R_{122}^{\phantom{000}1}
    -R_{133}^{\phantom{000}1}-R_{144}^{\phantom{000}1}
    -R_{233}^{\phantom{000}2}-R_{244}^{\phantom{000}2}
    -R_{344}^{\phantom{000}3}.
\end{equation}
It now becomes apparent that several of these components are
undefined at the throat if we allow the shape function to have
the form $b=b(r,l)$: since $R$ is a scalar invariant, we have
a curvature singularity at the throat.  We conclude that $b$
must be a function of $r$ alone; since we now have
$\partial b/\partial l\equiv 0$, the singularity disappears.
So we return to $b=b(r)$ and $b(r_0)=r_0$.  It is interesting
to note that the redshift function can retain the proposed
form $\Phi=\Phi(r,l)$.

\section{The null energy condition (NEC)}
Let us recall from the Introduction that the NEC states
that for the energy-momentum tensor $T_{\alpha\beta}$,
$T_{\alpha\beta}k^{\alpha}k^{\beta}\ge 0$ for all null
vectors $k^{\alpha}$.  We also recall that an ordinary
Morris-Thorne wormhole can only be held open if this
condition is violated, thereby requiring exotic matter.
As noted earlier, our goal is to show that thanks to the
extra spatial dimension, the wormhole throat can be
threaded with ordinary matter, but there is a violation
of the NEC in the fifth dimension.

To that end, we start with the four-dimensional null vector
$(1,1,0,0)$.  Continuing with our  orthonormal frame, consider
the Einstein field equations
\begin{equation}
   G_{\hat{\alpha}\hat{\beta}}=R_{\hat{\alpha}\hat{\beta}}-\frac{1}
{2}Rg_{\hat{\alpha}\hat{\beta}}=8\pi T_{\hat{\alpha}\hat{\beta}},
\end{equation}
where
\begin{equation}\label{E:metrictensor}
   g_{\hat{\alpha}\hat{\beta}}=
   \left(
   \begin{matrix}
   -1&0&0&0\\
   \phantom{-}0&1&0&0\\
   \phantom{-}0&0&1&0\\
   \phantom{-}0&0&0&1
   \end{matrix}
   \right).
\end{equation}
As before, $T_{00}=\rho$ is the energy density and
$T_{11}=p_r$ is the radial pressure, but instead of
$\rho+p_r$, we need to consider the more general form
\begin{multline}\label{E:exotic}
   8\pi T_{\alpha\beta}k^{\alpha}k^{\beta}=G_{00}
   +G_{11}+2G_{01}=\left[R_{00}-\frac{1}{2}R(-1)\right]+\\
   \left[R_{11}-\frac{1}{2}R(1)\right]
   +2\left[R_{01}-\frac{1}{2}R(0)\right]
   =R_{00}+R_{11}+2R_{01}.
\end{multline}
Since we are primarily interested in the vicinity of the
throat, we recall that $1-b(r_0)/r_0=0$.  So by Eq.
(\ref{E:R01}), $R_{01}=0$ at the throat and we are back
to the original $\rho +p_r$.  So from Eqs. (\ref{E:R00})
and (\ref{E:R11}), we have (recalling that $\partial b/
\partial l\equiv 0$)
\begin{multline}\label{E:NEC}
   8\pi(\rho +p_r)|_{r=r_0}=\frac{b'(r_0)-1}{r_0^2}\
   \left[1+\frac{r_0}{2}\frac{\partial\mu(r_0,l,t)}{\partial r}
   \right]+\frac{\partial^2\Phi(r_0,l)}{\partial l^2}
   e^{-2\mu(r_0,l,t)}\\-\frac{\partial\Phi(r_0,l)}{\partial l}
   \frac{\partial\mu(r_0,l,t)}{\partial l}e^{-2\mu(r_0,l,t)}
   +\left(\frac{\partial\Phi(r_0,l)}{\partial l}\right)^2
   e^{-2\mu(r_0,l,t)}\\-e^{-2\Phi(r_0,l)}
   \frac{\partial^2\mu(r_0,l,t)}{\partial t^2}
   -e^{-2\Phi(r_0,l)}\left(\frac{\partial\mu(r_0,l,t)}
   {\partial t}\right)^2.
\end{multline}

To make the analysis tractable, let us assume for now that
our wormhole is time independent; so it is convenient to use
the notation $\mu =\mu(r_0,l)$.  If, in addition, $\Phi$ is
independent of $l$, then Eq. (\ref{E:NEC}) becomes
\begin{equation}
  8\pi(\rho +p_r)|_{r=r_0}=\frac{b'(r_0)-1}{r_0^2}
  \left[1+\frac{r_0}{2}\frac{\partial\mu(r_0,l)}{\partial r}
  \right],
\end{equation}
which is the condition discussed in Ref. \cite{pK18a}, and leads
to the conclusion (since $b'(r_0)<1$) that
\begin{equation*}
  \rho+p_r>0 \quad  \text{at} \quad r=r_0
\end{equation*}
provided that
\begin{equation}\label{E:condition1}
   \frac{\partial\mu(r_0,l)}{\partial r}<-\frac{2}{r_0}.
\end{equation}
The $l$-dependent terms in Eq. ({\ref{E:NEC}) could strengthen
the conclusion if
\begin{equation}\label{E:ldependent}
   \frac{\partial^2\Phi(r_0,l)}{\partial l^2}>0 \quad
   \text{and} \quad \frac{\partial\Phi(r_0,l)}{\partial l}
   <0.
   \end{equation}
(We will return to this point in the next section.)   
Since Condition  (\ref{E:ldependent}) depends on $l$, 
the conclusion goes well beyond Ref. \cite{pK18a}.  
Now, since $\Phi$ is a function of both $r$ and $l$,
let us return to Inequality (\ref{E:condition1}) and
concentrate on the null vector $(1,0,0,0,1)$ in the
five-dimensional space:
\begin{multline}
   8\pi T_{\alpha\beta}k^{\alpha}k^{\beta}=G_{00}
   +G_{44}+2G_{04}=\left[R_{00}-\frac{1}{2}R(-1)\right]+\\
   \left[R_{44}-\frac{1}{2}R(1)\right]
   +2\left[R_{04}-\frac{1}{2}R(0)\right]
   =R_{00}+R_{44}+2R_{04}.
\end{multline}
While $g_{04}=0$ from Eq. (\ref{E:metrictensor}), observe
that by Eq. (\ref{E:R04}), $R_{04}$ is also equal to zero
since $\partial b/\partial l\equiv 0$.  Still assuming time
independence, the condition $1-b(r_0)/r_0=0$ then yields
 \begin{equation}\label{E:condition2}
   8\pi (\rho +p_r)|_{r=r_0}=
    \frac{1}{2}\frac{b'(r_0)-1}{r_0}\left[
    -\frac{\partial\Phi(r_0,l)}{\partial r}
    +\frac{\partial\mu(r_0,l)}{\partial r}\right].
\end{equation}
Inequality (\ref{E:condition1}),
$\partial\mu(r_0,l)/\partial r<-2/r_0$, implies that the
second factor on the right-hand side of
Eq. (\ref{E:condition2}) is positive if
\begin{equation}\label{E:condition3}
   \frac{\partial\Phi(r_0,l)}{\partial r}=-A<
   \frac{\partial\mu(r_0,l)}{\partial r}<-\frac{2}{r_0},
\end{equation}
strikingly similar to Inequality (\ref{E:condition1}).
Since $b'(r_0)<1$, it follows from Eq. (\ref{E:condition2})
that $\rho(r_0)+p_r(r_0)<0$ in the fifth dimension.  So
the NEC is indeed violated even though the throat of the
wormhole is threaded with ordinary matter.

\emph{Remark:} One physical consequence of the NEC is
that it forces the local energy density to be positive.
Also, in the four-dimensional case, the NEC must be
satisfied for all null vectors if the throat is to be
threaded with ordinary matter.  That these requirements
can be met has already been shown in Ref. \cite{pK18a} and
need not be repeated here.

\section{The size of the extra dimension and \\the time-dependence}
Returning to line element (\ref{E:L4}), it is interesting to note
that apart from the exponential functions, $\mu$ itself does not
appear as a factor in the components of the Riemann curvature
tensor, but only its derivatives.  This allows $\mu$ to have
virtually any magnitude.  So if $\mu(r,l,t)$ is negative and
large in absolute value, then $e^{2\mu(r,l,t)}$ is necessarily
small and may even be curled up.  (The same condition on $\mu$ 
would increase $8\pi(\rho +p_r)|_{r=r_0}$ in Eq. (\ref{E:NEC}) 
on account of Condition (\ref{E:ldependent}) and thereby 
strengthen the conclusion.)

Turning now to the time dependence, observe that in Eq. (\ref{E:NEC}),
the time-dependent terms in the expression for $8\pi (\rho +p_r)$
can be written in the form
\begin{equation}\label{E:H}
   H=-e^{-2\Phi(r_0,l)}\left[\frac{\partial^2\mu(r_0,l,t)}{\partial t^2}
   +\left(\frac{\partial\mu(r_0,l,t)}{\partial t}\right)^2\right].
\end{equation}
Suppose that $|\mu(r,l,t)|$ is relatively small.  Then Eq. (\ref{E:H})
may lead to an interesting interpretation: being confined to a small
range of values and changing with time, $\mu(r,l,t)$ could undergo a
slow oscillation.  Then $H$ in Eq. (\ref{E:H}) is positive near any
peak, where $\partial^2\mu(r,l,t)/\partial t^2<0$ and
$\partial\mu(r,l,t)/\partial t$ is close to zero.  This behavior
could greatly increase the value of $H$ during certain periods.
So judging from Eq. (\ref{E:NEC}), there will be periods in
which $\rho +p_r>0$ at $r=r_0$ even if Inequality
(\ref{E:condition1}) is weakened.

\section{Conclusions}
An earlier paper by the author \cite{pK18a} assumes that $\Phi$,
$b$, and $\mu$ are functions of the radial coordinate $r$ only.
In this paper we extend this model by assuming that $\Phi$,
$b$, and $\mu$ are functions of both $r$ and $l$ and that, in
addition, $\mu$ is time dependent.  It is shown that an
unrestricted dependence on the fifth dimension could lead
to a curvature singularity at the throat, forcing $b$ to be
a function of $r$ only.  The main goal is to determine  the
conditions under which the NEC is met in the four-dimensional
setting, thereby allowing the throat of the wormhole to be
threaded with ordinary matter.  The same conditions lead to
a violation of the NEC in the fifth dimension.  This
violation is the key factor in sustaining the wormhole structure.

The dependence of $\Phi$ on $l$ could favorably affect the
ability to use ordinary matter near the throat.  The same is
true for the dependence of $\mu$ on $t$ during certain periods.
Finally, there is no restriction on the size of $\mu$.  So if
$\mu$ is negative and large in absolute value, the extra
dimension would be extremely small and may even be curled up.
\\
\\
\textbf{Conflicts of Interest}
\\
\\
\noindent
The author declares that there are no
conflicts of interest regarding the
publication of this paper.


\begin{thebibliography}{9}
\bibitem{MT88}Morris, M.S. and Thorne, K.S. (1988) Wormholes in
   Spacetime and Their Use for Interstellar Travel: A Tool for
   Teaching General Relativity. \emph{American Journal of Physics},
    \textbf{56}, 395-412.
\bibitem{pK09}Kuhfittig, P.K.F. (2009) Theoretical Construction of
   Morris-Thorne Wormholes Compatible with Quantum Field Theory.
    arXiv: 0908.4233.
\bibitem{sS05}Sushkov, S.V. (2005) Wormholes Supported by a
   Phantom Energy. \emph{Physical Review D}, \textbf{71}, ID:
   043520.
\bibitem{fL05}Lobo, F.S.N. (2005) Phantom Energy Traversable
   Wormholes. \emph{Physical Review D}, \textbf{71}, ID: 084011.
\bibitem{pK18a}Kuhfittig, P.K.F. (2018) Traversable Wormholes
   Sustained by an Extra Spatial Dimension. \emph{Physical Review
    D}, \textbf{98}, ID: 064041.
\bibitem{pK18}Kuhfittig, P.K.F. (2018) Two Diverse Models of
   Embedding Class One. \emph{Annals of Physics}, \textbf{392},
   63-70.
\bibitem{LO09}Lobo, F.S.N. and Oliveira, M.A. (2009) Wormhole
   Geometries in $f(R)$ Modified Theories of Gravity.
   \emph{Physical Review D}, \textbf{80}, ID: 104012.
\bibitem{fR12}Rahaman, F., Kuhfittig, P.K.F., Ray, S. and Islam, S.
   (2012) Searching for Higher Dimensional Wormholes with
   Noncommutative Geometry. \emph{Physical Review D}, \textbf{86},
   ID: 106010.
\bibitem{pK15}Kuhfittig, P.K.F. (2015) Macroscopic Traversable
   Wormholes with Zero Tidal Forces Inspired by Noncommutative
   Geometry. \emph{International Journal of Modern Physics D},
    \textbf{24}, ID: 1550023.
\bibitem{KS96}Kar, S. and Sahdev, D. (1996) Evolving Lorentzian
   Wormholes. \emph{Physical Review D}, \textbf{53}, 722-730.
\bibitem{CBA96}Cataldo, M., Bahamonde, S. and Arostica, F. (2013)
   $(N+1)$-dimensional Lorentzian Evolving Wormholes Supported by
   Polytropic Matter. \emph{European Physical Journal C}, \textbf{73},
   ID: 2517.
\bibitem{pW11}Wesson, P.S. (2011) The Cosmological `Constant' and
   Quantization in Five Dimensions. \emph{Physics Letters B},
   \textbf{706}, 1-5.
\bibitem{WP92}Wesson P.S. and Ponce de Le\'{o}n, J. (1992)
   Kaluza-Klein Equations, Einstein's Equations, and an Effective
   Energy-Momentum Tensor. \emph{Journal of Mathematical Physics},
   \textbf{33}, 3883-3887.
\bibitem{pW15}Wesson, P.S. (2015) The Status of Modern
   Five-Dimensional Gravity. \emph{International Journal of  Modern
   Physics D }, \textbf{24}, ID: 1530001.
\bibitem{pW13}Wesson, P.S. (2013) Astronomy and the Fifth Dimension.
   arXiv: 1301.0033.
\bibitem{HT90}Hughes, L.P. and Tod, K.P. An Introduction to General
   Relativity (Cambridge University Press, Cambridge, England,
   1990).

\end{thebibliography}
\end{document}